\documentclass{aa}
\usepackage[varg]{txfonts}
\usepackage[separate-uncertainty=true]{siunitx}
\usepackage[version=3]{mhchem}

\def\aap{A\&A}

\def\apjs{ApJS}
\def\apjl{ApJL}

\begin{document}

\title{NIR spectroscopy of the Sun and HD20010}
\subtitle{Compiling a new linelist in the NIR}

\author{ D.~T.~Andreasen\inst{1,2}
    \and S.~G.~Sousa\inst{1}
    \and E.~Delgado Mena\inst{1}
    \and N.~C.~Santos\inst{1,2}
    \and M.~Tsantaki\inst{1}
    \and B.~Rojas-Ayala\inst{1}
    \and V.~Neves\inst{3}}

\institute{
Instituto de Astrof\'isica e Ci\^encias do Espa\c{c}o, Universidade do Porto, CAUP, Rua das Estrelas, 4150-762 Porto, Portugal
\email{daniel.andreasen@astro.up.pt}
\and
Departamento de F\'isica e Astronomia, Faculdade de Ci\^encias, Universidade do Porto, Rua Campo Alegre, 4169-007 Porto, Portugal
\and
Departamento de F\'{i}sica, Universidade Federal do Rio Grande do Norte, 59072-970 Natal, RN, Brazil
}

\date{Received ...; accepted ...}

\abstract
{Effective temperature, surface gravity, and metallicity are basic
spectroscopic stellar parameters necessary to characterize
a star or a planetary system. Reliable atmospheric parameters for
FGK stars have been obtained mostly from methods that relay on high
resolution and high signal-to-noise optical spectroscopy. The
advent of a new generation of high resolution near-IR spectrographs
opens the possibility of using classic spectroscopic methods with
high resolution and high signal-to-noise in the NIR spectral window.}
{We aim to compile a new iron line list in the NIR from a solar
spectrum to derive precise stellar atmospheric parameters,
comparable to the ones already obtained from high resolution optical
spectra. The spectral range covers \SI{10000}{\angstrom} to
\SI{25000}{\angstrom}, which is equivalent to the Y, J, H, and K bands.}
{Our spectroscopic analysis is based on the iron excitation and
ionization balance done in LTE. We
use a high resolution and high signal-to-noise ratio spectrum of the Sun
from the Kitt Peak telescope as a starting point to compile the iron
line list. The oscillator strengths ($\log\mathit{gf}$) of the iron lines were calibrated for the Sun.
The abundance analysis was done using
the MOOG code after measuring equivalent widths of 357 solar iron lines.}
{We successfully derived stellar atmospheric parameters for the
Sun.
Furthermore, we analysed
HD20010, a F8IV star, from which we derived stellar atmospheric
parameters using the same line list as for the Sun. The spectrum
was obtained from the CRIRES-POP database.
The results are compatible with the ones found in the literature,
confirming the reliability of our line list. However, due to the
quality of the data we obtain large errors.}
{}

\keywords{data reduction: high resolution spectra --
          stars individual: HD20010 --
          stars individual: Sun}
\maketitle

\section{Introduction}
\label{sec:introduction}

Effective temperature ($T_\mathrm{eff}$), surface gravity ($\log g$),
and metallicity ([M/H], where iron is normally used as a proxy)
are fundamental atmospheric parameters necessary to characterise a single
star, as well as to determine other indirect fundamental parameters,
such as mass, radius, and age from stellar evolutionary models
\citep[see e.g.][]{Girardi2000,Dotter2008,Baraffe2015}.
Precise and accurate stellar parameters are also essential in
exoplanet searches. Planetary radius and mass are mainly found from
lightcurve analysis and radial velocity analysis, respectively. The
determination of the mass of the planet implies a knowledge of the
stellar mass, while the measurement of the radius of the planet
is dependent on our capability to derive the radius of the star
\citep[see e.g.][]{Torres2008,Ammler2009,Torres2012}.

The derivation of precise stellar atmospheric parameters is not a simple
task. Different approaches often lead to discrepant results \citep[see
e.g.][]{Santos13}. Interferometry is usually considered as an accurate
method to derive stellar radii \citep[e.g.][]{Boyajian2012}, however,
is only applicable for bright nearby stars. Asteroseismology, on the
other hand, reveals the inner stellar structure by observing the stellar
pulsations at the surface. From asteroseismology it is possible to
measure the surface gravity and mean density, and therefore calculate
the mass and radius \citep[e.g.][]{Kjeldsen1995}.

A key parameter for the indirect determination of stellar bulk
properties is the effective temperature. In that respect, the infrared
flux method (IRFM) have proven to be reliable for FGK dwarf and
subgiant stars. However, the IRFM needs a priori knowledge of the
bolometric flux, reddening, surface gravity, and stellar metallicity
\citep{Blackwell1977,Ramirez2005b,Casagrande2010}.

Finally, the use of high resolution spectroscopy along with stellar
atmospheric models is an extensively tested method that allows
the derivation of the fundamental parameters of a star \citep[see
e.g.][]{Santos13,Valenti2005}. The procedure depends on the quality
of the spectra, their resolution, and wavelength region. For low
resolution spectra ($\lambda/\Delta\lambda < 20\,000$) it is
preferred to fit the overall observed spectrum with a synthetic one
\citep[see e.g.][]{Recio2006}. For higher resolution spectra of
slowly rotating stars (below 10 to 15 \si{km/s}) we are in the regime
where the equivalent width (EW) method can be used (for details see
Sect.~\ref{sec:method}).

The derivation of stellar atmospheric parameters from high
resolution spectra in the optical is now based on a standard
procedure \citep[see e.g.][]{Valenti2005,Sousa2008a}. With the
advancement of high resolution NIR instruments, we will now be able to use a similar
technique as used in the optical part of the spectrum
\citep[see e.g.][]{Melendez1999,Sousa2008a,Tsantaki2013,Mucciarelli2013,Bensby2014}.
At the moment, the GIANO spectrograph installed at TNG is already
available \citep{GIANO}, as well as the IRD spectrograph installed
at Subaru \citep{IRD}. Three new spectrographs are planned for the
near future: 1) CARMENES for the \SI{3.5}{m} telescope at Calar Alto
Observatory \citep{CARMENES} had first light at December 2014,
2) CRIRES+ at VLT \citep{CRIRESp} with expected first light in 2017,
and 3) SPIRou at CFHT \citep{SPIROU1,SPIROU2} with expected first light
in 2017 as well. The spectral resolutions for these spectrographs range
between $50\,000$ and $100\,000$.

Even though reliable line lists for the derivation of stellar parameters
using optical spectra exist, the situation is different in the near-IR
regime. There exists a few, e.g. \citet{Onehag2012,Origlia2013,Rhodin2015},
for the synthesis method and the large general compilation by
\citet{Melendez1999}. Moreover, there are line lists compiled
in the NIR for the APOGEE survey \citep[see e.g.][]{Smith2013,Shetrone2015}.
Thus, in this paper we want to explore the
possibility to create a line list of iron lines in the NIR which can
be applied for FGKM stars optimized for the EW method in a consistent
way as it is currently done for these stars in the optical (FGK
only). The paper is organized as follows: In Sect.~\ref{sec:method}
we present how to compile a line list and the method for deriving
parameters with the equivalent width method for an iron line list.
In Sect.~\ref{sec:results} we present the results for the derived
parameters for the Sun and HD20010. Lastly, we discuss our results in
Sect.~\ref{sec:conclusion}.

\section{Method}
\label{sec:method}

The two most widely used methods for deriving stellar atmosphere
parameters from a spectrum are spectral synthesis and the EW method.
The spectral synthesis method compares synthetic spectra to an observed
spectrum and finds the best model by a minimization procedure
\citep[see e.g.][]{Valenti2005,Onehag2012,Blanco2014}. When the minimization
procedure reaches a minimum, the final atmospheric parameters are found.

The equivalent width (EW) method
\citep[see e.g.][]{Sousa2008a,Mucciarelli2013,Bensby2014}, which we use
in this work, is based on the measurements of EWs from a list of lines
combined with the matching atomic data. The EW for a single line is
given as:
\begin{align}
    \label{eq:EW}
    EW = \int_0^\infty \left(1 - \frac{F_\lambda}{F_0}\right) d\lambda,
\end{align}
where $F_0$ is the continuum level and $F_\lambda$ is the flux as a
function of wavelength.

Using this method, we obtain the abundance of individual lines by
the radiative transfer code MOOG \citep[][version 2013]{Sneden1973}
under the assumption of local thermodynamic equilibrium (LTE). To
obtain metallicity, we expect every spectral line of the same element
to produce the same abundance. In our analysis, we use neutral iron
(\ion{Fe}{i}) and single ionized iron (\ion{Fe}{ii}) as a proxy for
the metallicity. The effective temperature and surface gravity are
derived from the principles of ionization and excitation equilibrium
\citep[see][]{Gray2006}.

A disadvantage of the EW method, may be a miscalculation of the EW. This
can have the source in a misplacement of the continuum level, which
leads to and over- or underestimation of the EW for the given line.
Another source of error is contamination with either telluric lines or
other neighbouring lines. The relative error is typically larger for the
weak lines. In this work we will focus at the spectral region covered by
the Y, J, H, and K bands, which covers more than $\SI{15000}{\AA}$.

\subsection{Compiling the line list}

To compile the line list we used the VALD3 database \citep{VALD1,VALD2}.
First, we downloaded a list of all iron lines present in the near
infrared region, covering $10\,000\si{\AA}$ to $25\,000\si{\AA}$.
In total, $78\,537$ iron lines were found in this spectral region
($50\,198$ \ion{Fe}{i} lines and $28\,339$ \ion{Fe}{ii} lines).
Many of these lines are too faint to be detected in a spectrum
of a solar type star. A spectrum of the Sun was downloaded from
the BASS2000 web page\footnote{The web page can be found here:
\url{bass2000.obspm.fr/solar_spect.php}} to select the best lines
for this analysis. The NIR part of the spectrum were obtained from
the Kitt Peak telescope \citep{Hinkle1995} at a resolution of
\SI{0.004}{\angstrom} at \SI{10000}{\angstrom} to \SI{0.1}{\angstrom}
at \SI{50000}{\angstrom}. The spectrum was downloaded in the highest
possible resolution at a given wavelength. The signal-to-noise ratio
of the spectrum varies from 3000 at $\SI{12000}{\AA}$ down to 1400 at
$\SI{21400}{\AA}$.

We use the ARES software\footnote{The ARES software can be found here:
\url{http://www.astro.up.pt/~sousasag/ares/}. The following settings
were used: lambdai=7500, lambdaf=54000, smoothder=4, space=2.0,
rejt=0.995, lineresol=0.07, and miniline=2.}\citep{Sousa2007,Sousa2015a}
to automatically measure EWs of all the lines. Since the first version
of ARES expect a 1D spectrum with equidistant wavelength spacing,
the solar spectrum was interpolated to a regular grid with constant
wavelength step of \SI{0.01}{\angstrom}. This did not change the
appearance of the spectrum, and hence not the EW. The EWs are measured
by fitting Gaussian profiles to spectral lines. For a given line, ARES
outputs the central wavelength of the line, the number of lines fitted
for the final result, the depth of the line, the FWHM of the line, the EW
of the line, and Gaussians coefficients for the line.

Once this step was done we then selected a subset of lines using the
following criteria:
\begin{itemize}
    \item If the number of fitted lines by ARES for a given line is higher than 10,
        this line is rejected because it is believed to be severely blended.
    \item If the EW is lower than \SI{5}{m\angstrom} for an absorption line, the strength
        is too low and it may be difficult to see the line in spectra with low
        signal-to-noise ratio or a spectrum with many spectral features.
    \item If the EW is higher than \SI{200}{m\angstrom} for a given line, the strength
        is too high and we can no longer fit the line with a Gaussian profile, since
        the absorption line no longer has a pure Gaussian profile.
    \item If the fitted central wavelength is more than $\SI{0.05}{\AA}$ away
        from the wavelength provided by VALD3, the line will also be rejected to
        avoid false identification.
\end{itemize}
After the automatic removal of lines following the above criteria
we reduced the number of lines to 6060 and 2735 for \ion{Fe}{i} and
\ion{Fe}{ii}, respectively.

\subsection{Visual removal of lines}
\label{sub:visual_removal_of_lines}

A visual inspection of the lines is necessary at this point in order
to allow us to select only the best lines. The best lines we define as
lines which are not blended, and therefore reliable EW measurements can
be made.

In this step we analyzed in detail small \SI{3}{\angstrom} wide spectral
windows around each line. For each spectral window, the corresponding
absorption lines for all elements were downloaded from the VALD3
database. The location of these lines were plotted on top of the solar
spectrum, and any iron line was excluded if a line of another element
was present at the same wavelength. Iron lines were also excluded when
the absorption line was severely blended by other spectral lines. Many
of the removed iron lines at this step have high excitation potential,
compared to the final line list, since these lines are generally weaker
than those with lower excitation potential. After this step we were down
to 593 \ion{Fe}{i} lines and 22 \ion{Fe}{ii} lines.

For some spectral regions it was not clear which element or elements
caused an absorption line. In these cases the iron lines were marked for
further investigation with synthesis explained below.

\subsection{Synthesis of selected lines}
\label{sub:synthesis_of_selected_lines}

\begin{figure}[tpb]
    \centering
    \includegraphics[width=1.0\linewidth]{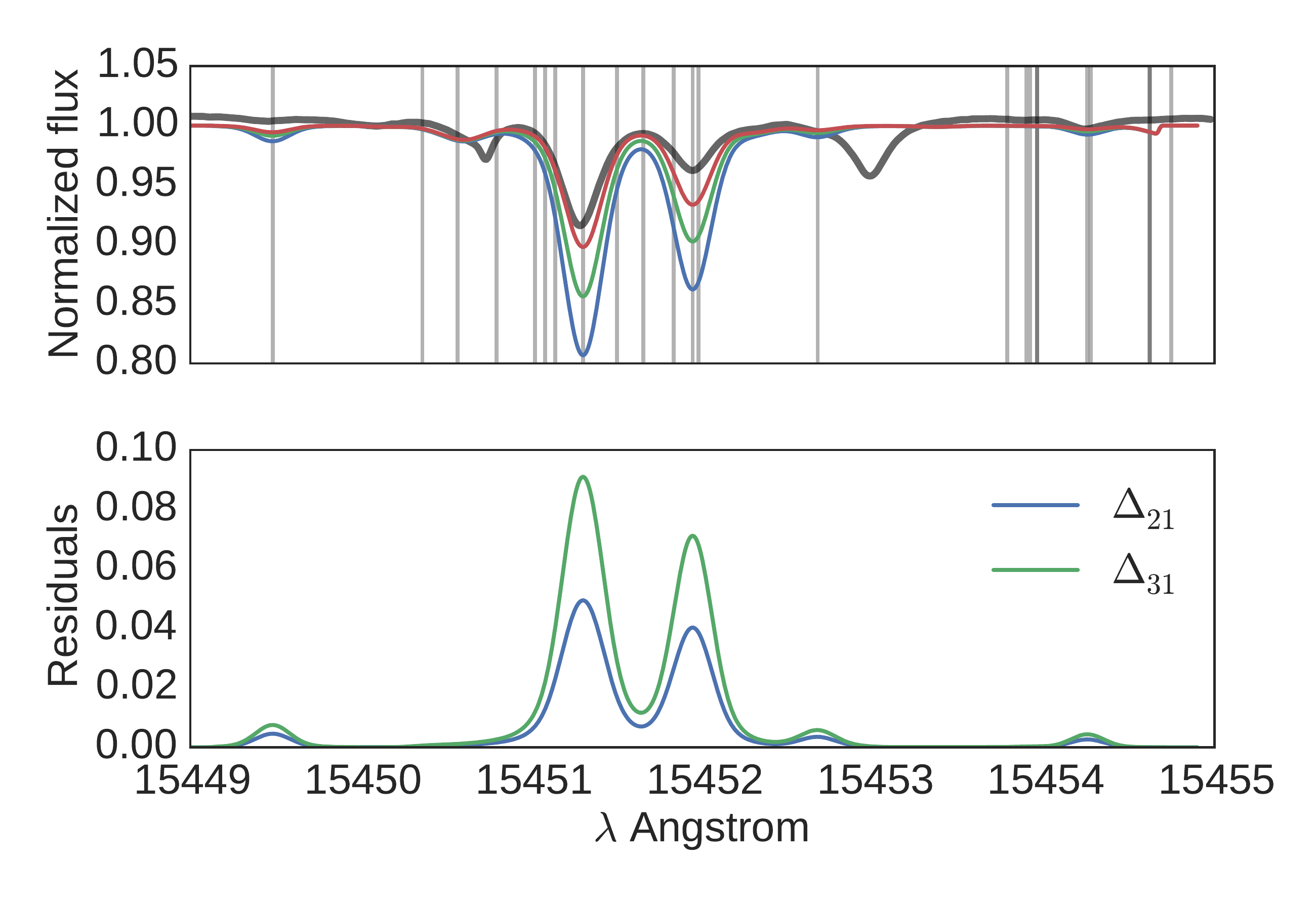}
    \caption{Top panel shows the observed spectra in grey, while the
    colored graphs is synthetic spectra with increasing iron abundance
    as the central two lines get deeper. The iron abundance is varied
    0.4 dex in total. The vertical lines show all the places, where there are
    iron lines in the line list. Bottom panel shows two curves, which is
    the difference between the first synthetic spectrum and the second,
    $\Delta_{21}$, and the difference between the first synthetic
    spectrum and the third, $\Delta_{31}$. This is for highlighting
    where the change in iron abundance has an impact.}
    \label{fig:synthesis}
\end{figure}

Lines from all elements in a $\SI{6}{\AA}$ window around an iron line
marked for further investigation were used to make a synthetic spectrum.
The synthetic spectra were made with MOOG with the \emph{synth} driver.
We use an ATLAS9 atmosphere model \citep{Kurucz1993} with the following
nominal solar atmospheric parameters: $T_\mathrm{eff}=\SI{5777}{K}$,
$\log g = 4.438$, and $\xi_\mathrm{micro} = \SI{1.0}{km/s}$ to resemble
the Sun. We used 3 different iron abundances for the synthesis. One with
solar iron abundance, the second with 0.2 dex above solar and the third
with 0.2 dex below solar. We consider a solar iron abundance of 7.47 as
presented in \cite{Gonzalez2000}. This choice of solar parameters and
iron abundances was done to match the ones already used by our team in
previous papers \citep[see e.g.][and references therein]{Santos13} and
thereby provide consistency within our group. If the synthetic spectra
shows variation at the absorption line of interest with respect to the
different iron abundances, then it is likely to be an iron line. We also
changed abundances of other elements in the proximity to see if our line
is blended with other elements. An example of these plots can be seen in
Fig~\ref{fig:synthesis}.

Sometimes more than one iron line might be present with very similar
wavelengths so they can no longer be resolved. In order to find the
iron line which is creating the observed absorption line, one of the
two were excluded from the line list for the synthetic spectra. If this
removed (either fully or partially) the absorption line in the synthetic
spectra, then it will be the cause for the observed absorption line,
otherwise we excluded the line from the line list presented in this
work.

A few times two iron lines had identical wavelengths and excitation
potential. In those cases the $\log \mathit{gf}$ were combined (sum of
the $\mathit{gf}$-value) to create a single line that can be analyzed
with our method. We ended up with 414 and 12 lines of \ion{Fe}{i} and
\ion{Fe}{ii}, respectively.

\subsection{Calibrating the line list: astrophysical $\log$ gf values}
\label{ssub:Recalibrating-the-atomic-data}

The iron abundances for each line were calculated using the same
solar atmosphere model as described above for synthesis. This step
allowed us to remove possible outliers based on the assumption that
errors in the $\log \mathit{gf}$ values from the VALD3 database
would never lead to variations of the derived iron abundance of more
than 1 dex. All the \ion{Fe}{i} lines before recalibration of
the oscillator strength and removal of lines which deviates more than
1 dex are presented in Fig.~\ref{fig:fe1_before_recal}.
Note that we only removed \ion{Fe}{i} lines here, since
the \ion{Fe}{ii} lines are sparse and essential to determine the
surface gravity when we reach ionization balance, as explained in
Sect.~\ref{sec:deriving_parameters_with_the_ew_method}. After removal
of 1 dex outliers we are down to 319 and 12 lines, for \ion{Fe}{i} and
\ion{Fe}{ii} respectively.

\begin{figure}[tpb]
    \centering
    \includegraphics[width=1.0\linewidth]{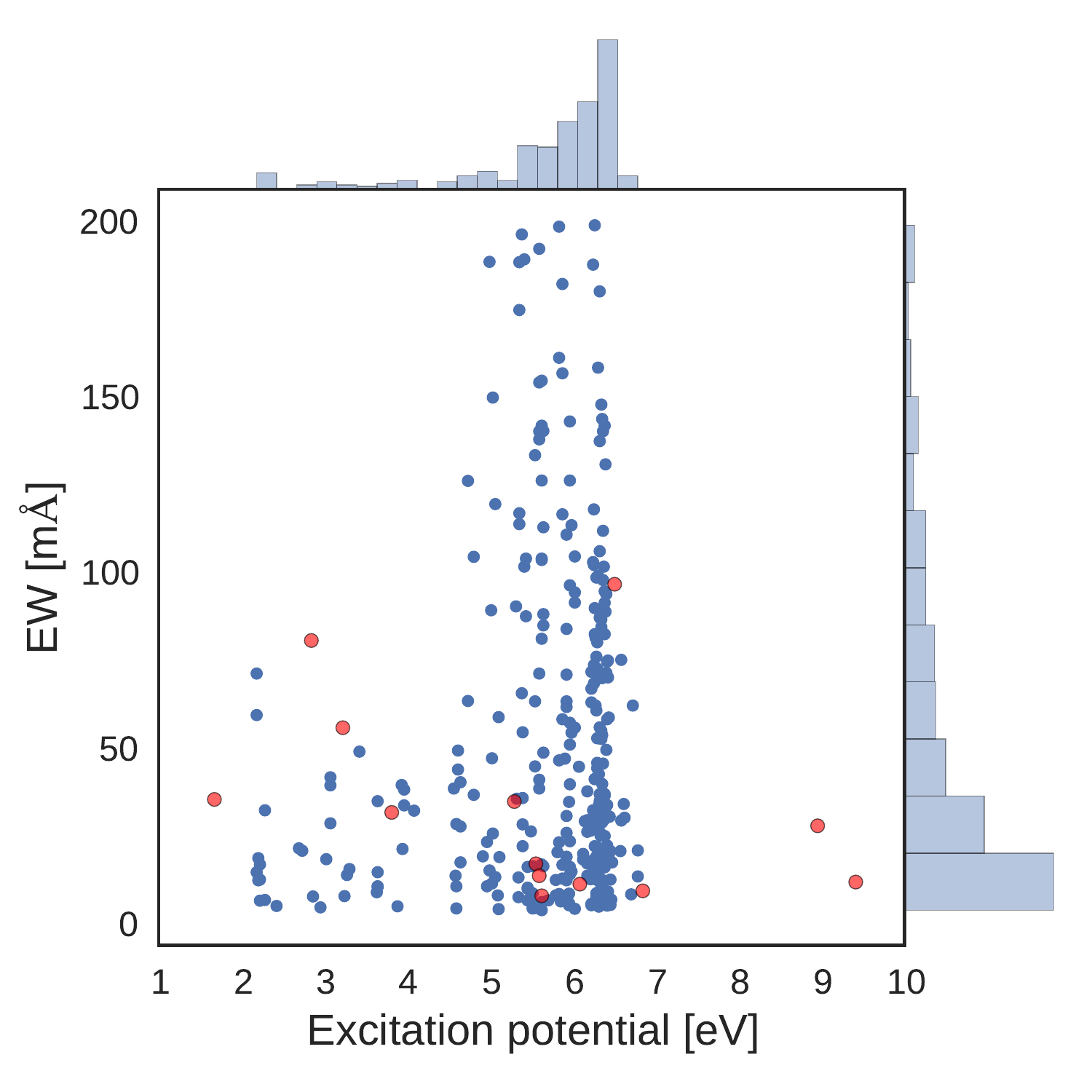}
    \caption{The distribution of \ion{Fe}{i} and \ion{Fe}{ii} lines,
    colored blue and red, respectively. The distribution shows the
    measured EWs for the Sun as a function of the excitation potential.}
    \label{fig:EWvsEP}
\end{figure}

\begin{figure}[tpb]
    \centering
    \includegraphics[width=1.05\linewidth]{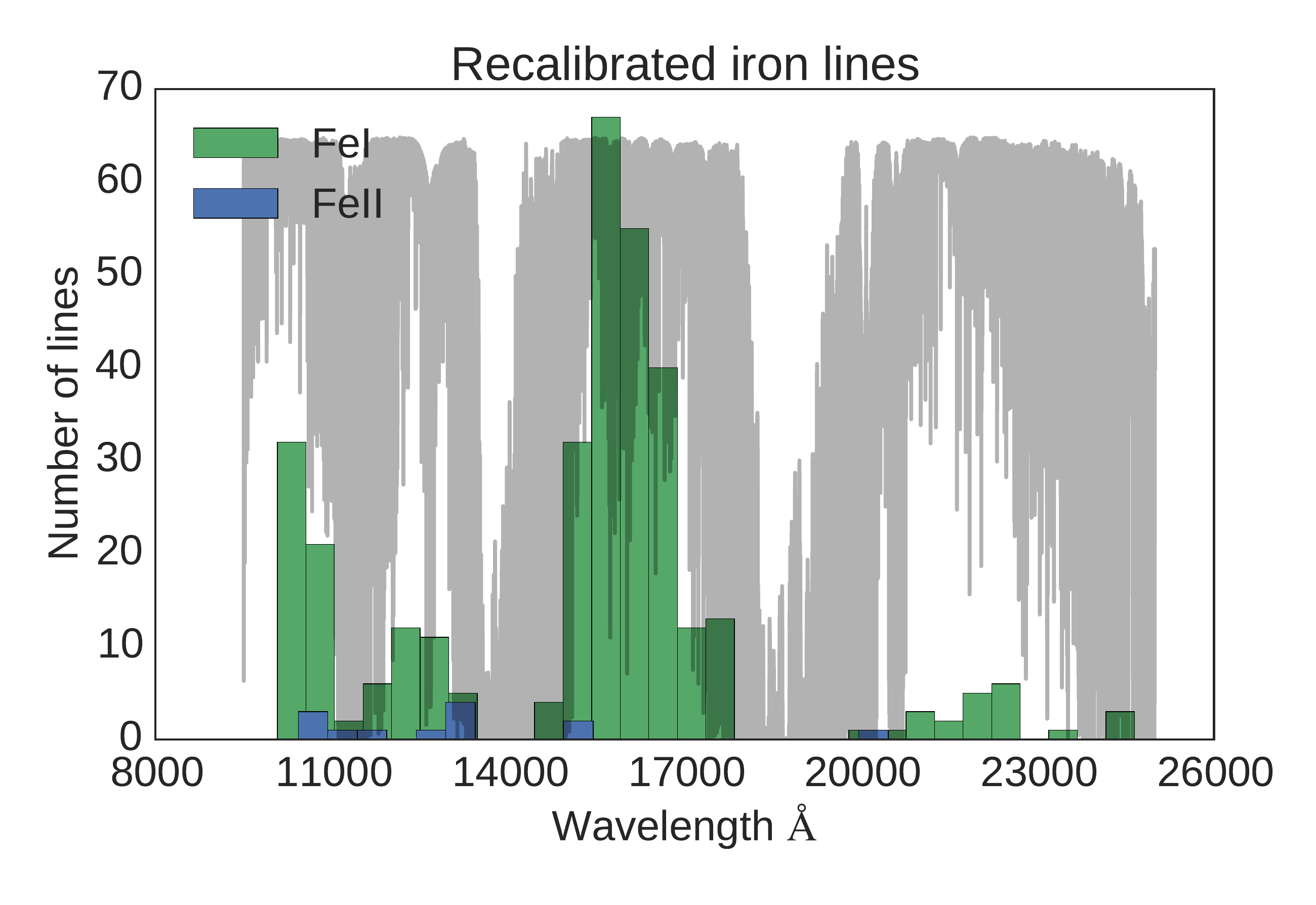}
    \caption{Distribution of both \ion{Fe}{i} and \ion{Fe}{ii} lines
    on top of the solar spectrum. The distributions are for the final
    line list. There are two areas in the spectrum with high telluric
    contamination, which also mark the border between the filters we
    use: from J to H around 14000\si{\angstrom} and from H to K around
    19000 \si{\angstrom}. Most of the lines are located in the H band.}
    \label{fig:solarspectrum}
\end{figure}

After the removal of lines from the complete VALD3 line list we
recalibrate the oscillator strength of the lines ($\log\mathit{gf}$) in
order to match the adopted solar abundance, an inverse solar analysis.
This allows to perform a differential analysis for other stars. Similar
approaches have been done by \citet{Sousa2008a,Onehag2012,Rhodin2015}.
In Fig.~\ref{fig:EWvsEP} the EWs of the iron lines present in the Sun
are plotted as a function of the excitation potential. This plot shows
the distribution after recalibration of $\log \mathit{gf}$ after the
cut for lines with abundances deviating more than 1 dex from the solar
value. The majority of the iron lines are found in H band as shown in
Fig~\ref{fig:solarspectrum}.

\subsection{Removal of high dispersion lines}
\label{sub:removal_of_unstable_lines}

\begin{figure}[tbp!]
    \centering
    \includegraphics[width=1.0\linewidth]{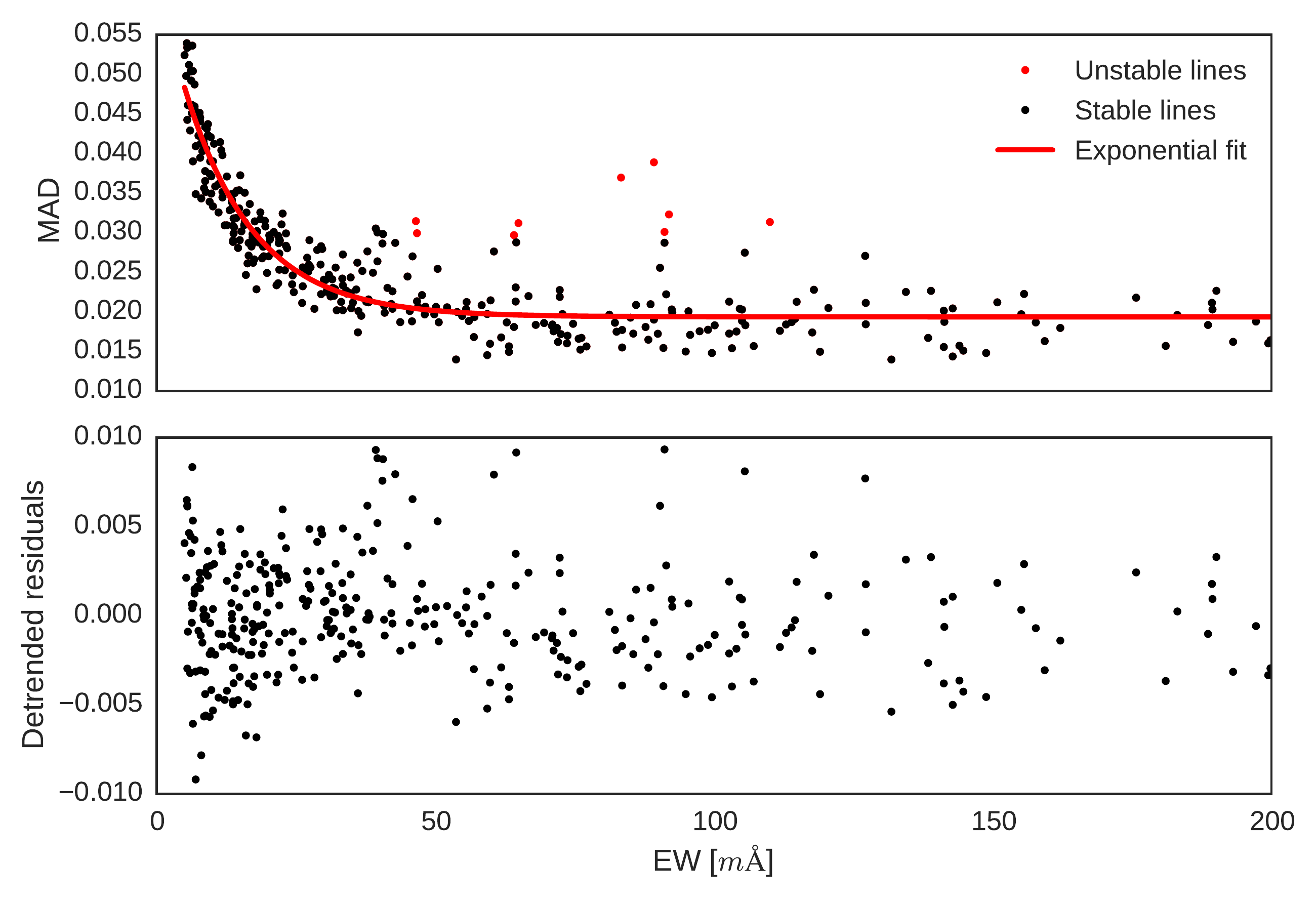}
    \caption{The upper plot shows the MAD values of 100 noisy lines
    with a simulated signal-to-noise ratio of 50. The red points are
    the 23 detected unstable lines, and the red curve is the last fit
    in the iterative removal of unstable lines. The lower plot show the
    detrended points from the upper plot, from where the $3\sigma$ is
    calculated for each iteration.}
    \label{fig:unstable_lines}
\end{figure}

To chose which line-derived abundances are less prone to errors caused
by the uncertainties in the EW measured, we decided to do the following
test. A Gaussian distribution is made for the EW of each line. We use
the width for the Gaussian distribution following the formula presented
in \citet{Caryel1988} below:
\begin{align}
    \sigma \simeq 1.6 \frac{\sqrt{\Delta\lambda\; \mathrm{EW}}}{\mathrm{S/N}},
\end{align}
where $\Delta\lambda=0.1\si{\angstrom}$ and we consider a
signal-to-noise ratio of 50, much lower than the signal-to-noise ratio
of the spectrum. This width is used to create a Gaussian distribution
with a mean around the original EW.
\begin{align}
    f(x, EW, \sigma) = \frac{1}{\sqrt{2\pi\sigma^2}} e^{-\frac{(x-EW)^2}{2\sigma^2}}.
\end{align}
We make 100 draws for each line and derive the abundance with solar
parameters, using the same atmospheric model as described above.
For each line we calculate the mean absolute deviation (MAD). The
MAD values are plotted against the original EWs in the upper part
of Fig.~\ref{fig:unstable_lines}. We see a clear trend towards
weaker lines, which is expected since a small absolute change in
the EW result in a large relative change in abundance, hence a high
MAD value. However, this does not mean the abundances of these
lines have a high dispersion. Therefore, we detrend the data with
a fitted exponential. The residuals are shown in the lower part
of Fig.~\ref{fig:unstable_lines}. We use the detrended data as a
measurement for the dispersion of a given line. A single point above $3
\sigma$ is then removed iteratively until there are no longer any points
above this threshold. In this process we remove 33 lines. The final line
list is presented in Table~\ref{tab:linelist}.

\begin{table}[tb!]
    \caption{The final line list as found for the Sun with astrophysical
    $\log\mathit{gf}$ values. A complete version of this table is will be available
    online.}
    \label{tab:linelist}
    \centering
    \begin{tabular}{ccrrr}
      \hline\hline
        Wavelength [$\si{\angstrom}$] & Element      & EP [$\si{eV}$] & $\log\mathit{gf}$ &  EW [$\si{m\angstrom}$]   \\
      \hline
        10070.521                     & \ion{Fe}{i}  &     5.51       &      -1.527       &   6.6 \\
        10080.415                     & \ion{Fe}{i}  &     5.10       &      -2.008       &   5.3 \\
        10081.394                     & \ion{Fe}{i}  &     2.42       &      -4.551       &   6.2 \\
        10137.100                     & \ion{Fe}{i}  &     5.09       &      -1.768       &   9.2 \\
        10142.844                     & \ion{Fe}{i}  &     5.06       &      -1.574       &  14.4 \\
        10155.163                     & \ion{Fe}{i}  &     2.18       &      -4.340       &  15.8 \\
        10156.507                     & \ion{Fe}{i}  &     4.59       &      -2.125       &  11.8 \\
        10167.469                     & \ion{Fe}{i}  &     2.20       &      -4.199       &  19.8 \\
        10195.106                     & \ion{Fe}{i}  &     2.73       &      -3.625       &  21.9 \\
        10227.991                     & \ion{Fe}{i}  &     6.12       &      -0.449       &  19.4 \\
        10230.796                     & \ion{Fe}{i}  &     6.12       &      -0.414       &  21.0 \\
        10265.218                     & \ion{Fe}{i}  &     2.22       &      -4.668       &   7.7 \\
        10327.340                     & \ion{Fe}{i}  &     5.54       &       0.504       & 134.4 \\
        10332.328                     & \ion{Fe}{i}  &     3.63       &      -3.145       &  10.1 \\
        10340.886                     & \ion{Fe}{i}  &     2.20       &      -3.672       &  46.7 \\
        10347.966                     & \ion{Fe}{i}  &     5.39       &      -0.754       &  36.9 \\
        10353.805                     & \ion{Fe}{i}  &     5.39       &      -1.035       &  23.2 \\
        10364.063                     & \ion{Fe}{i}  &     5.45       &      -1.129       &  17.3 \\
        10379.000                     & \ion{Fe}{i}  &     2.22       &      -4.246       &  18.0 \\
        10388.746                     & \ion{Fe}{i}  &     5.45       &      -1.527       &   7.8 \\
          \ldots                      &   \ldots     &    \ldots      &      \ldots       &  \ldots\\
        10427.305                     & \ion{Fe}{ii} &     6.08       &      -1.662       &  12.4 \\
        10501.498                     & \ion{Fe}{ii} &     5.55       &      -1.926       &  18.2 \\
        10862.644                     & \ion{Fe}{ii} &     5.59       &      -2.043       &  14.8 \\
        11125.580                     & \ion{Fe}{ii} &     5.62       &      -2.301       &   9.1 \\
        11833.056                     & \ion{Fe}{ii} &     2.84       &      -3.379       &  81.7 \\
        12913.876                     & \ion{Fe}{ii} &     6.50       &       0.045       &  97.7 \\
        13251.144                     & \ion{Fe}{ii} &     9.41       &       0.860       &  13.0 \\
        13277.306                     & \ion{Fe}{ii} &     5.29       &      -2.043       &  35.9 \\
        13294.853                     & \ion{Fe}{ii} &     3.22       &      -3.613       &  56.9 \\
        13419.109                     & \ion{Fe}{ii} &     3.81       &      -3.484       &  32.8 \\
        15247.133                     & \ion{Fe}{ii} &     6.84       &      -1.691       &  10.5 \\
        15350.156                     & \ion{Fe}{ii} &     8.95       &       0.602       &  29.0 \\
        20460.070                     & \ion{Fe}{ii} &     1.67       &      -5.758       &  36.5 \\
      \hline
    \end{tabular}
\end{table}

\subsection{Deriving parameters with the EW method}
\label{sec:deriving_parameters_with_the_ew_method}

Once the EWs have been measured for all iron lines in the line list
(or as many as possible), the next step is to derive the atmospheric
parameters. Atmosphere models are necessary for computing abundances of
the lines. The literature offers the possibility to choose from a wide
variety of model atmospheres. Models like ATLAS9 \citep{Kurucz1993} and
MARCS \citep{Gustafson2008} have been the preferred atmosphere models
for derivation of spectroscopic parameters for FGK stars.

We use the ATLAS9 models which, for efficiency, are created in a grid
according to effective temperature, surface gravity, and metallicity.
In order to search for final parameters it is necessary to interpolate
models from the grid, thus allowing to look into a finer grid space
\citep[see e.g.][]{Sousa2014}. This grid of atmosphere models have been
used extensively by our group which allow us to work consistently over
multiple wavelength regions (optical and NIR).

For a given atmosphere model, abundances of all the lines in the line
list are calculated. By removing any correlation between the excitation
potential and abundance of all lines (from same element) the effective
temperature is constrained. In a similar way, the microturbulence
can be constrained by removing any correlation between the reduced EW
($\log EW/\lambda$) and iron abundances, and the surface gravity is
found when there is ionization balance, i.e. the mean abundance of
\ion{Fe}{i} and \ion{Fe}{ii} are equal. Lastly, the iron abundance comes
from calculating the mean of all the iron abundances.

When there is no longer any correlation, the final atmospheric
parameters are obtained from the last atmosphere model.

In order to find the best atmosphere model, a minimization algorithm
is used based on the downhill simplex method \citep{Press1992} which
searches in the parameter space for the best fitting atmospheric model,
i.e. the best parameters. The convergence criteria for the correlation
between excitation potential and abundances is a slope lower than
0.001. A slope lower than 0.002 for the correlation between the reduced
EW and the abundances, and a difference of less than 0.005 between
the mean abundances for \ion{Fe}{i} and \ion{Fe}{ii} as used in e.g.
\cite{Sousa2008a} and \cite{Tsantaki2013}.

The error estimate is based on the same method presented in
\citet{Gonzalez1998}. The uncertainty in $\xi_\mathrm{micro}$ is
determined from the standard deviation in the slope of abundance
versus reduced equivalent width, the uncertainty of $T_\mathrm{eff}$
is determined from the uncertainty in the slope of abundance
versus excitation potential in addition to the uncertainty in
$\xi_\mathrm{micro}$, the uncertainty in the iron abundance
is a combination of the uncertainties in $T_\mathrm{eff}$,
$\xi_\mathrm{micro}$ and the scatter of the individual \ion{Fe}{i}
abundances. The uncertainty in the surface gravity is based on the
uncertainty in $T_\mathrm{eff}$ and the scatter in \ion{Fe}{ii}
abundances.

\section{Results}
\label{sec:results}

\subsection{Derived parameters for the Sun}
\label{sec:derived_parameters_of_the_sun}

We derived the stellar atmospheric parameters for the Sun using the
resulting line list (including the solar calibrated astrophysical $\log
\mathit{gf}$ values). We used the minimization procedure described in
Sect.~\ref{sec:deriving_parameters_with_the_ew_method}. Since the line
list and $\log\mathit{gf}$ values have been selected using the solar
spectrum, it is with no surprise that the derived parameters for the Sun
perfectly match the adopted solar values within the error bars as seen
in Table~\ref{tab:solar_params}.

\begin{figure}[tbp!]
    \centering
    \includegraphics[width=1.0\linewidth]{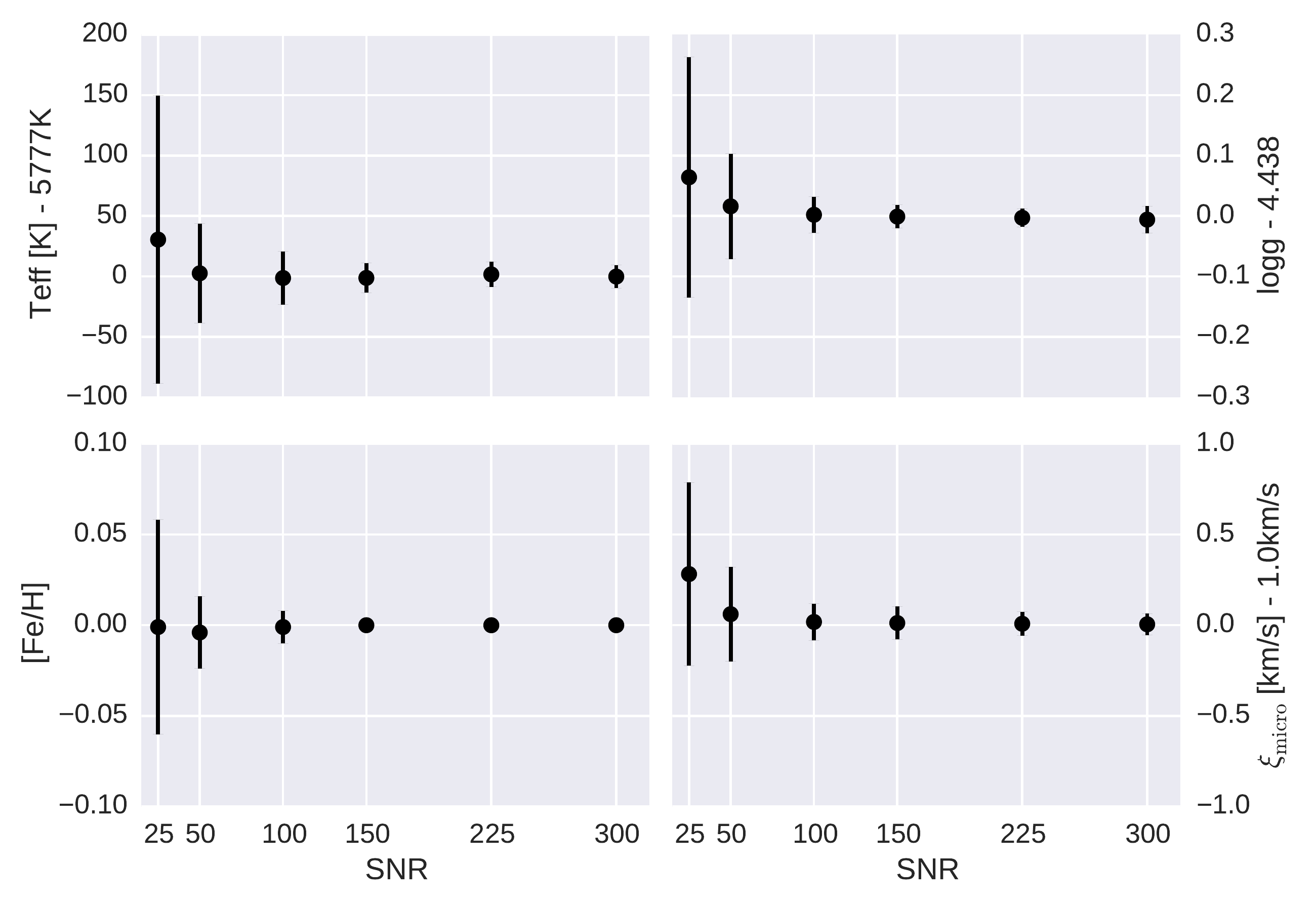}
    \caption{All plots shows derived parameters as a function of the
    signal-to-noise. The error is the 3$\sigma$ standard deviation from
    the 10 different runs for each signal-to-noise. The upper left plot
    shows the effective temperature. The upper right plot show the
    surface gravity ($\log g$). The lower left shows the iron abundance,
    used as a proxy for the metallicity. Finally, in the lower right
    plot the microturbulence is shown.}
    \label{fig:snr_sun}
\end{figure}

Moreover, we derive parameters for different signal-to-noise ratios,
namely 25, 50, 100, 150, 225, and 300. The signal-to-noise ratios
are obtained by drawing EW from a Gaussian distribution with widths
dependent on the EW itself and the signal-to-noise as described above
in Sect.~\ref{sub:removal_of_unstable_lines}. For each considered
signal-to-noise, we make 10 random line lists, giving us a total of 60
line lists. This exercise shows the expected precision for different
signal-to-noise ratios with the proposed line list. The final results
are presented in Fig.\ref{fig:snr_sun}. The error bars represent the
$3\sigma$ standard deviation from the 10 different runs. As seen
from the figure, we expect to derive precise parameters (effective
temperature more precise than $\SI{50}{K}$, surface gravity with a
precision of 0.1 dex, iron abundance with a precision of 0.05, and
microturbulence with a precision of 0.3) down to a signal-to-noise ratio
of 50. At higher signal-to-noise ratios the precision increases. E.g.
at a signal-to-noise ratio at 100 the erros are reduced by a factor
of 2. The results can also be seen in Table~\ref{tab:solar_params}.
This shows that the line list is fully reliable for the whole range of
signal-to-noise ratios considered, even if the precision decreases at
lower signal-to-noise ratios as expected.

\begin{table*}[htb!]
    \caption{The derived parameters for the Sun at different
    signal-to-noise ratios. The error is the 3$\sigma$ standard
    deviation calculated from the 10 runs at each signal-to-noise
    ratio.}
    \label{tab:solar_params}
    \centering
    \begin{tabular}{lllll}
      \hline\hline
        SNR & $T_\mathrm{eff}$ (K) & $\log g$ (dex)  &  [Fe/H] (dex)    & $\xi_\mathrm{micro}$ (km/s)  \\
      \hline
  Original  &  $5776 \pm 0$        & $4.43 \pm 0.00$ & $0.00 \pm 0.00$  & $0.99 \pm 0.00$              \\
      \hline
        25  &  $5808 \pm 119$      & $4.50 \pm 0.20$ & $0.00 \pm 0.06$  & $1.28 \pm 0.51$              \\
        50  &  $5780 \pm 41$       & $4.45 \pm 0.09$ & $0.00 \pm 0.02$  & $1.06 \pm 0.26$              \\
       100  &  $5776 \pm 22$       & $4.44 \pm 0.03$ & $0.00 \pm 0.01$  & $1.02 \pm 0.10$              \\
       150  &  $5776 \pm 12$       & $4.44 \pm 0.02$ & $0.00 \pm 0.00$  & $1.01 \pm 0.09$              \\
       225  &  $5779 \pm 10$       & $4.44 \pm 0.01$ & $0.00 \pm 0.00$  & $1.01 \pm 0.07$              \\
       300  &  $5777 \pm 10$       & $4.43 \pm 0.02$ & $0.00 \pm 0.00$  & $1.01 \pm 0.06$              \\
      \hline
    \end{tabular}
\end{table*}

\subsection{Derived parameters for HD20010}
\label{sec:derived_parameters_of_hd20010}

For testing our new line list we search for a well studied solar type
star. The spectrum for such a target needs to be available in the NIR at
both high resolution and high signal-to-noise. An ideal place to look
for such a star is the CRIRES-POP database \citep{Lebzelter2012}. Here,
the best target for testing is HD20010, a F8 subgiant star. This star
has been part of many surveys and is therefore well studied. Different
parameters from the literature are listed in Table~\ref{tab:parameters}.

\begin{table*}[htb!]
    \caption{Selection of literature values for the atmospheric
    parameters for HD20010. The mean and a $3 \sigma$ standard
    deviation is presented at the end of the table from the literature
    values included, which we use as a reference for our
    derived parameters.}
    \label{tab:parameters}
    \centering
    \begin{tabular}{l|llll}
      \hline\hline
     Author                 & $T_\mathrm{eff}$ (K) & $\log g$ (dex)  &  [Fe/H] (dex)    & $\xi_\mathrm{micro}$ (km/s)  \\
      \hline
    \cite{Balachandran1990} & $6152$               & $4.15$          & $-0.27\pm0.08$   & $1.6$                        \\
    \cite{Favata1997}       & $6000$               & \ldots          & $-0.35\pm0.07$   & \ldots                       \\
    \cite{Santos2004}       & $6275\pm57$          & $4.40\pm0.37$   & $-0.19\pm0.06$   & $2.41\pm0.41$                \\
    \cite{Gonzalez2010}     & $6170\pm35$          & $3.93\pm0.02$   & $-0.206\pm0.025$ & $1.70\pm0.09$                \\
    \cite{Ramirez2012}      & $6073\pm78$          & $3.91\pm0.03$   & $-0.30\pm0.05$   & \ldots                       \\
    \cite{Mortier2013}      & $6114$               & \ldots          & $-0.19$          & \ldots                       \\
      \hline
      Mean                  & $6131\pm255$         & $4.01\pm0.60$   & $-0.23\pm0.14$   & $1.90\pm1.08$                \\
      \hline
    \end{tabular}
\end{table*}

\begin{figure*}[tbp!]
    \centering
    \includegraphics[width=1.0\linewidth]{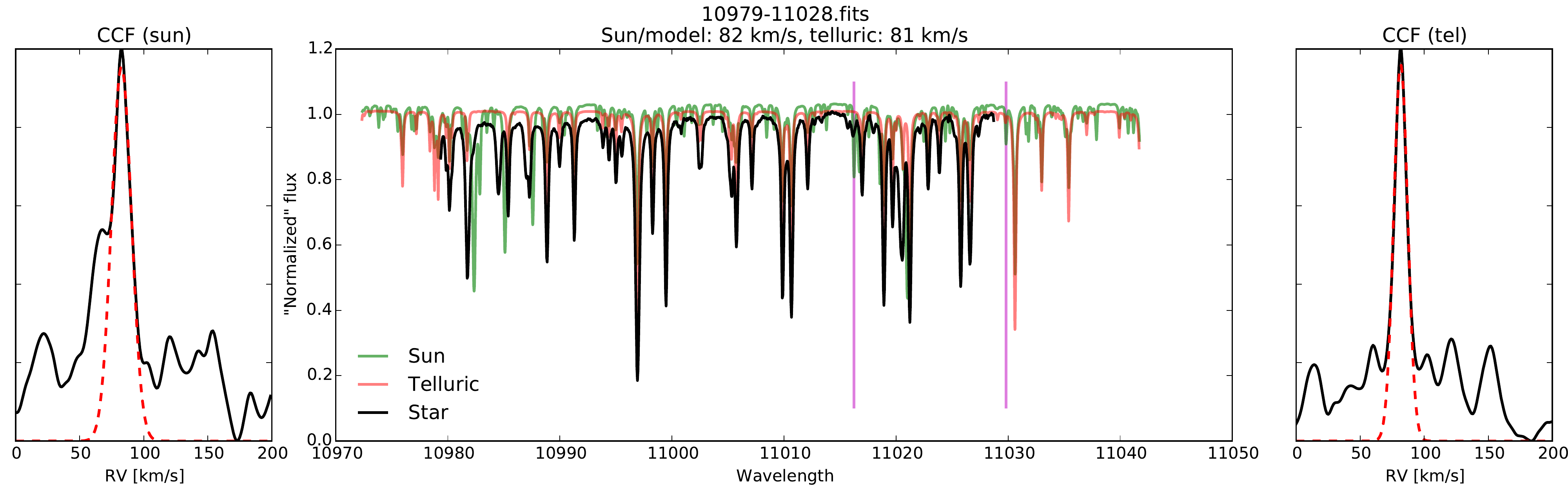}
    \caption{The middle plot shows a piece of HD20010 (black), the model
    spectrum, in this case the Sun (green), a telluric spectrum (red),
    and two lines from our line list (magenta vertical lines). The
    plot to the left shows the CCF of the Sun with a fitted Gaussian.
    The right plot shows the same as the one to the left, but for the
    telluric spectrum.}
    \label{fig:plot_fits}
\end{figure*}

The data available at CRIRES-POP are in the raw format and pipeline
reduced, while three small pieces of the spectra are fully reduced on the web
page\footnote{\url{http://www.univie.ac.at/crirespop/data.htm}}. The data
is in the standard CRIRES format with each fits file including four
binary tables with the data from the four detectors. In the future, the final
reduced data will be presented by the CRIRES-POP team. In contrast to the pipeline reduced data, this
will be of higher quality, a better wavelength calibration, and telluric correction. We
measured the EWs of the pipeline reduced spectra, and where there was
an overlap with the fully reduced spectrum, we measured both as a
consistency check. The measured EWs from the fully reduced spectra were
consistent with the measured EWs from the pipeline reduced spectra. As
mentioned above, we use the Y, J, H, and K bands which are all available
for this star. The spectra come in pieces of $\SI{50}{\angstrom}$
to $\SI{120}{\angstrom}$. These pieces have overlaps between each
other, and we were able to measure the EW for a single line up to
five times. Unfortunately, wavelength calibration is a difficult task
for CRIRES due to the rather small spectral regions measured on each
detector. Each calibration was performed separately for each detector
and required the availability of a sufficient number of calibration
lines in the respective spectral region. This was not always the case
and a default linear solution was applied. A pipeline reduced spectrum
shows up as a stretched spectrum if the wavelength calibration is poor
compared to e.g. a model spectrum or a solar spectrum. The wavelength
calibration does not have any effect of the signal-to-noise ratio, which
is generally high for the spectrum of HD20010. The signal-to-noise
varies between 200 and 400 for different chunks. The pipeline reduced
spectra for HD20010 contains tellurics and the wavelength is shifted in
radial velocity. All this combined make the line identification very
difficult. Therefore, we developed a software\footnote{The software
(plot\textunderscore{}fits) is open source and can be found here:
\url{https://github.com/DanielAndreasen/astro_scripts}} to proper
identify the lines. This software does the following:

\begin{enumerate}
    \item Plot the observed spectrum.
    \item Overplot a model spectrum. In this particular case the solar spectrum was
        used since the atmospheric parameters are close enough, so the sun can
        serve as a model.
    \item Overplot a telluric spectrum from the TAPAS web
          page\footnote{\url{http://ether.ipsl.jussieu.fr/tapas/}} \citep{Bertaux2014}.
    \item Overplot vertical lines at the location of lines in the list.
    \item Calculate the cross correlation function (CCF) for the telluric spectrum
        with respect to the observed spectrum, locate the maximum value by a Gaussian fit
        and use this to shift the telluric spectrum with the found RV.
    \item Do the same as the step above, but for the model.
    \item Shift the lines with the same RV as found for the model/solar spectrum.
\end{enumerate}
The final plot shows the shifted spectra, and the CCFs at the sides. An
example of the software in use is shown in Fig.~\ref{fig:plot_fits}. The
two RVs are part of the title of the plot.

Once the lines were identified, the EWs were measured with the
\emph{splot} routine in IRAF. The reason not to choose ARES for this
task was to visually confirm the identification of the line given the
relative poor wavelength calibration. We were able to measure 249
\ion{Fe}{i} lines and 5 \ion{Fe}{ii} lines compared to 344 \ion{Fe}{i}
lines and 13 \ion{Fe}{ii} lines for the Sun over the whole NIR spectral
region. Whenever we had more than one measurement of a line, the average
was used for the final EW.

We derived the stellar parameters using the standard procedure (see
Sect.~\ref{sec:deriving_parameters_with_the_ew_method}) as done for
the Sun. Given the relatively low quality of the spectrum of HD20010
(see below) and due to the fact that it is not corrected for telluric
contamination, we make a cut in EW at 5\si{m\angstrom}, in order to
remove the lines which are mostly affected by contamination from either
telluric or other line blends. Additionally, we make a cut in EP at
\SI{5.5}{eV}\footnote{We also derived the stellar parameters without
any cut in the EP, but the resulting values were always overestimated
(e.g., fixing $\log g$ to 4.01 we obtained a temperature of 6660K and
metallicity of +0.19 dex).}. We make this cut since the \ion{Fe}{i} and
\ion{Fe}{ii} lines usually used for stellar parameter determination
in the optical regime are also limited to similar values \citep[see
e.g.][]{Sousa2008a}. Higher excitation potential lines are also more
likely to be affected by non-LTE effects. When deriving the atmospheric
parameters, we make a $3\sigma$ outlier removal in the abundance
iteratively until there are no more outliers present. Since we could
only measure 5 \ion{Fe}{ii} lines, for comparison we also decided to
derive parameters using the same method, but fixing the surface gravity
to the reference value. The resulting atmospheric parameters
and iron abundances are presented in Table~\ref{tab:hd20010}. The
effective temperature, surface gravity, and metallicity agree within the
errors with the literature values. Similar parameters are obtained by
fixing $\log g$ to the average literature value or by letting it free.

\begin{table*}[htb!]
    \caption{The derived parameters for HD20010 with and without
    fixed surface gravity cut after 3$\sigma$ outlier removal.}
    \label{tab:hd20010}
    \centering
    \begin{tabular}{lllll}
      \hline\hline
                     & $T_\mathrm{eff}$ (K) &  $\log g$ (dex)  &   $\xi_\mathrm{micro}$ (km/s)  & [Fe/H] (dex)      \\
      \hline
        Literature   & $6131 \pm 255$       &  $4.01 \pm 0.60$ &    $1.90 \pm 1.08$              & $-0.23 \pm 0.14$ \\
      \hline
                     & $6116 \pm 224$       &  $4.21 \pm 0.58$ &    $2.45 \pm 0.45$              & $-0.14 \pm 0.14$ \\
                     & $6144 \pm 212$       &   4.01 (fixed)   &    $2.66 \pm 0.42$              & $-0.13 \pm 0.29$ \\
      \hline
    \end{tabular}
\end{table*}

\begin{figure}[tpb!]
    \centering
    \includegraphics[width=1.0\linewidth]{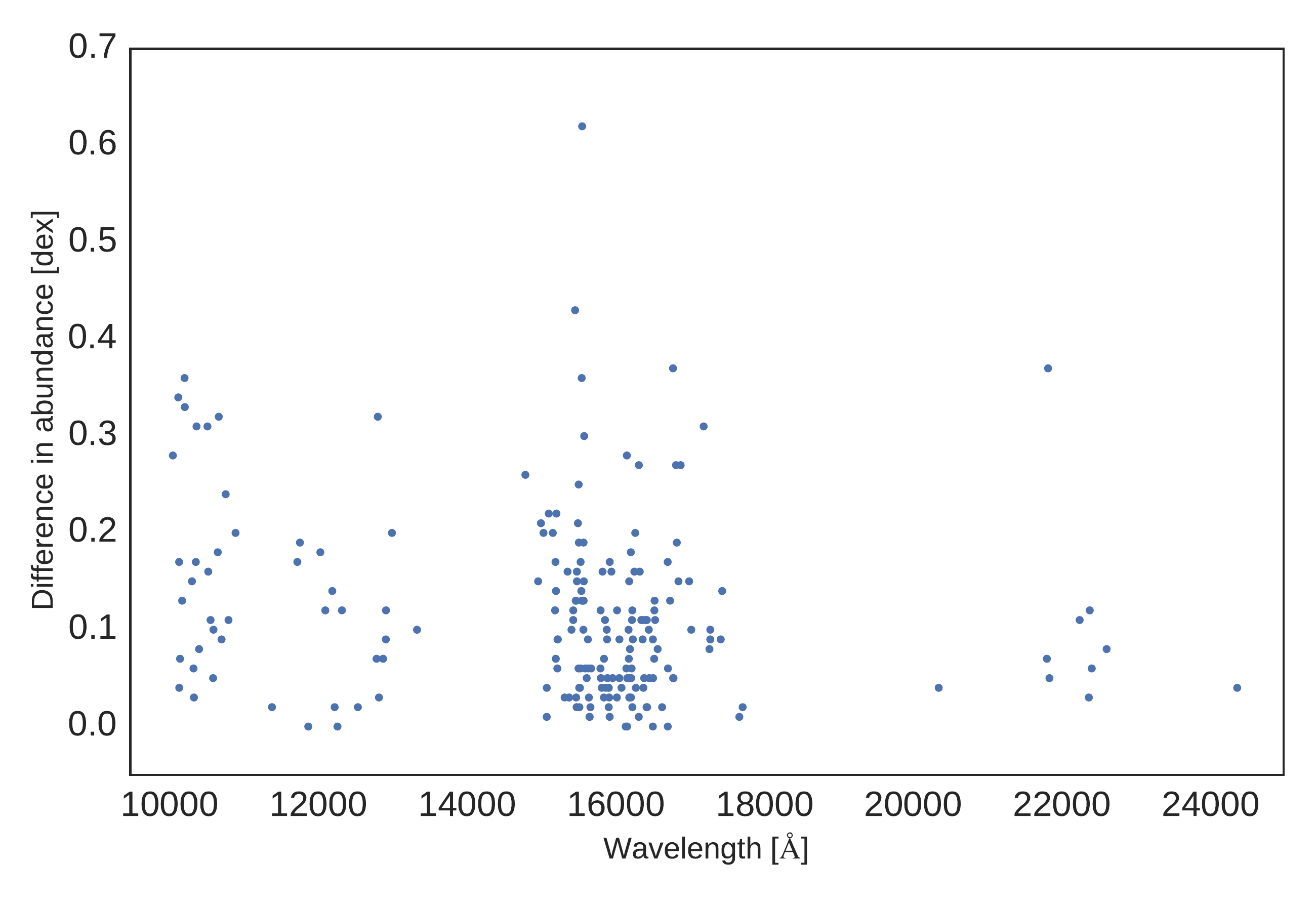}
    \caption{The difference in abundance for lines in HD20010 which
    have at least two measurements of EW. The difference is calculated
    between the highest measured EW and the lowest. For the line list
    used for HD20010 we used the mean of all measurements available.}
    \label{fig:abundance_error}
\end{figure}

The errors on the atmospheric parameters for HD20010 are quite
high compared to what is achievable with other measurements in the
literature, as presented above in Table~\ref{tab:parameters}. In order
to explain these errors, we calculate the abundances for all lines
which have at least two measurements of the EW. We then calculate the
abundances for the highest measured EW and the lowest. The differences
in abundances are presented in Fig.~\ref{fig:abundance_error}. The very
large differences (more than 0.1 dex) translates to the high errors in
the parameters.

\begin{figure}[tpb!]
    \centering
    \includegraphics[width=1.0\linewidth]{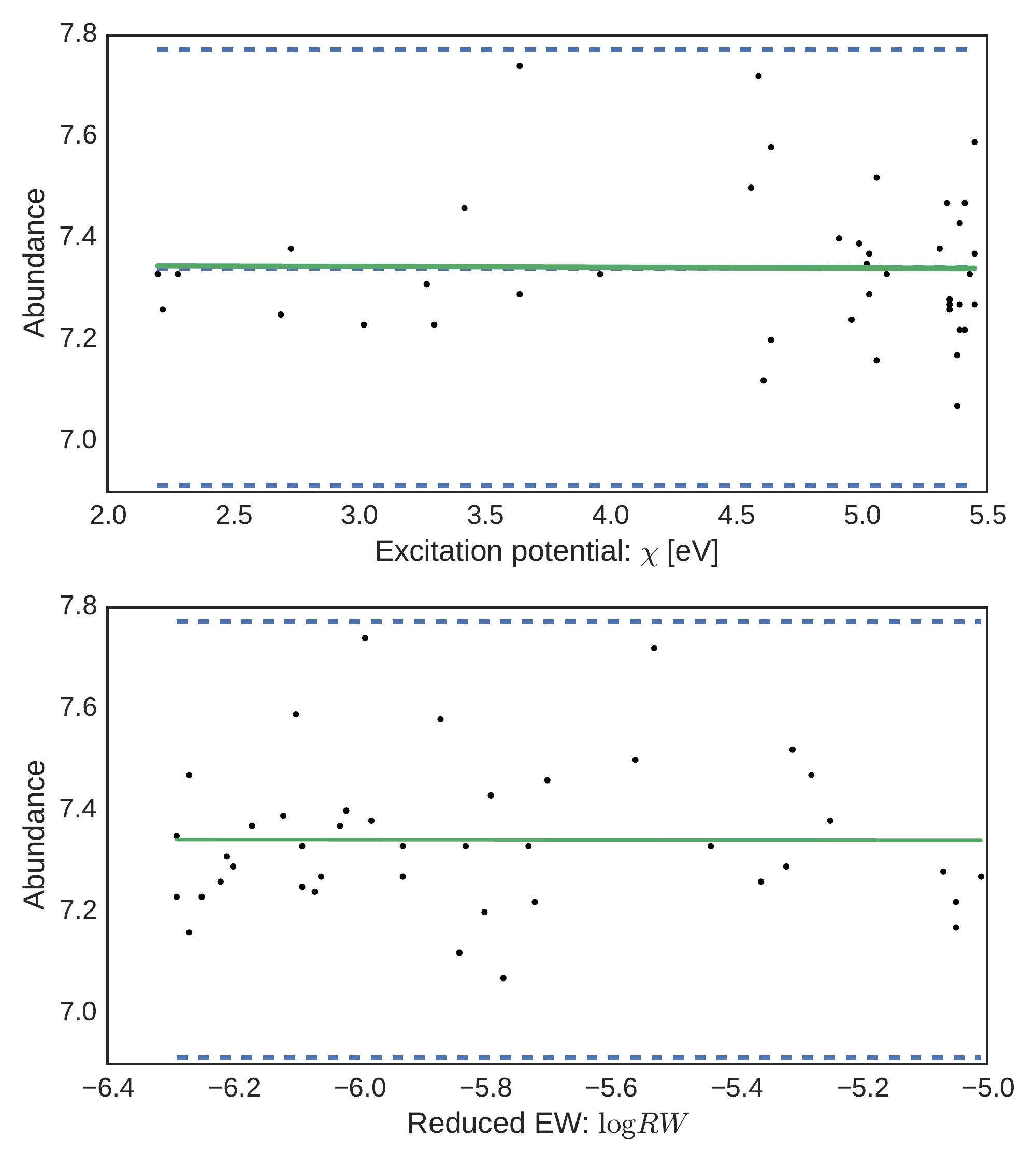}
    \caption{In the top plot the \ion{Fe}{i} abundances for all
    lines are shown as a function of EP. The bottom plot is again
    the \ion{Fe}{i} abundances, but against the reduced EW. The high
    dispersion in the abundances leads to high error bars on the derived
    atmospheric parameters. The green lines are the slopes, and the
    dashed lines are the mean (under the green line), and the $3 \sigma$
    standard deviation.}
    \label{fig:slopes}
\end{figure}

The source of the large errors on the parameters can be seen more
directly where abundances are compared to excitation potential or
abundances versus reduced EW. Here the dispersion on the abundances can
be seen directly, as shown in Fig.~\ref{fig:slopes}.

This test strongly suggest that errors in the EWs, likely due to the
poor quality of this spectrum, are responsible for the relatively large
error bars in the derived stellar parameters. Systematic errors (e.g.
due to a possible non-optimal reduction of the spectrum) may be the
reason for these large error bars. As the CRIRES-POP team continue their
great effort in reducing the spectra optimal, it will be interesting to
re-visit this star, once the entire spectrum is fully reduced.

\subsubsection{Surface gravity}
\label{subs:surface_gravity}
Although we have derived a consistent value for the surface gravity for
HD20010, given the small number of \ion{Fe}{ii} lines in the analysis,
this value should be considered with caution and of low precision. However,
we emphasize that from our experience in using this method (the ionization
balance) in the visible, the other atmospheric parameters ($T_\mathrm{eff}$,
and [Fe/H]) have a low interdependency with the surface gravity. This has
been shown by \citet{Torres2012} and more recently by \citet{Mortier2014}.
Furthermore, with the up-coming results from the \emph{Gaia} mission we will
get precise surface gravity for a large number of stars and thus the best
option would be to fix this parameter if necessary.

\section{Conclusion}
\label{sec:conclusion}

In this work, we present a new iron line list for the NIR. The quality
of the line list plays a key role for deriving atmospheric stellar
parameters. While the line list was compiled from a solar spectrum and
calibrated for the same, we tested it extensively for the slightly
hotter star, HD20010. The first results with this line list are
promising. We also show that for a spectrum that contain telluric
lines, the best results appear when removing lines with an EW lower
than $\SI{5}{m\angstrom}$. In the future, the development of new high
resolution NIR spectrographs will allow us to obtain more high quality
spectra of stars in the whole FGK spectral range, thus allowing us to
better test and refine this line list.

Furthermore, it will be interesting to explore the use of this line list
to derive parameters for M-dwarf stars using high resolution and high
signal-to-noise NIR spectra. M-dwarf stars are especially interesting
targets for an exoplanetary viewpoint, since they are prone to form
low mass exoplanets \citep{Bonfils2013}. Hence, a precise analysis
of the host star's atmospheric parameters may greatly improve our
characterization of the possible exoplanets orbiting these low mass
stars.

Lastly, with the up-coming NIR spectrographs as discussed above,
this work and future continuation will help the community to derive
atmospheric stellar parameters.

\begin{acknowledgements}

This work was supported by Funda\c{c}\~ao para a Ci\^encia e a
Tecnologia (FCT) through the research grants UID/FIS/04434/2013 and
PTDC/FIS-AST/1526/2014. N.C.S., and S.G.S. acknowledge the support from
FCT through Investigador FCT contracts of reference IF/00169/2012, and
IF/00028/2014, respectively, and POPH/FSE (EC) by FEDER funding through
the program “Programa Operacional de Factores de Competitividade
- COMPETE”. E.D.M. and B.J.A. acknowledge the support from FCT in
form of the fellowship SFRH/BPD/76606/2011 and SFRH/BPD/87776/2012,
respectively. This work also benefit from the collaboration of a
cooperation project FCT/CAPES - 2014/2015 (FCT Proc 4.4.1.00 CAPES).

This research has made use of the SIMBAD database operated at CDS,
Strasbourg (France).

This work has made use of the VALD database, operated at Uppsala
University, the Institute of Astronomy RAS in Moscow, and the University
of Vienna.

\end{acknowledgements}

\bibpunct{(}{)}{;}{a}{}{,}
\bibliographystyle{aa}
\bibliography{thesis}

\begin{appendix}
\section{Iron abundances before recalibrated $\log$ gf}
\label{sec:section label}

\begin{figure*}[tbp!]
    \centering
    \includegraphics[width=1.0\linewidth]{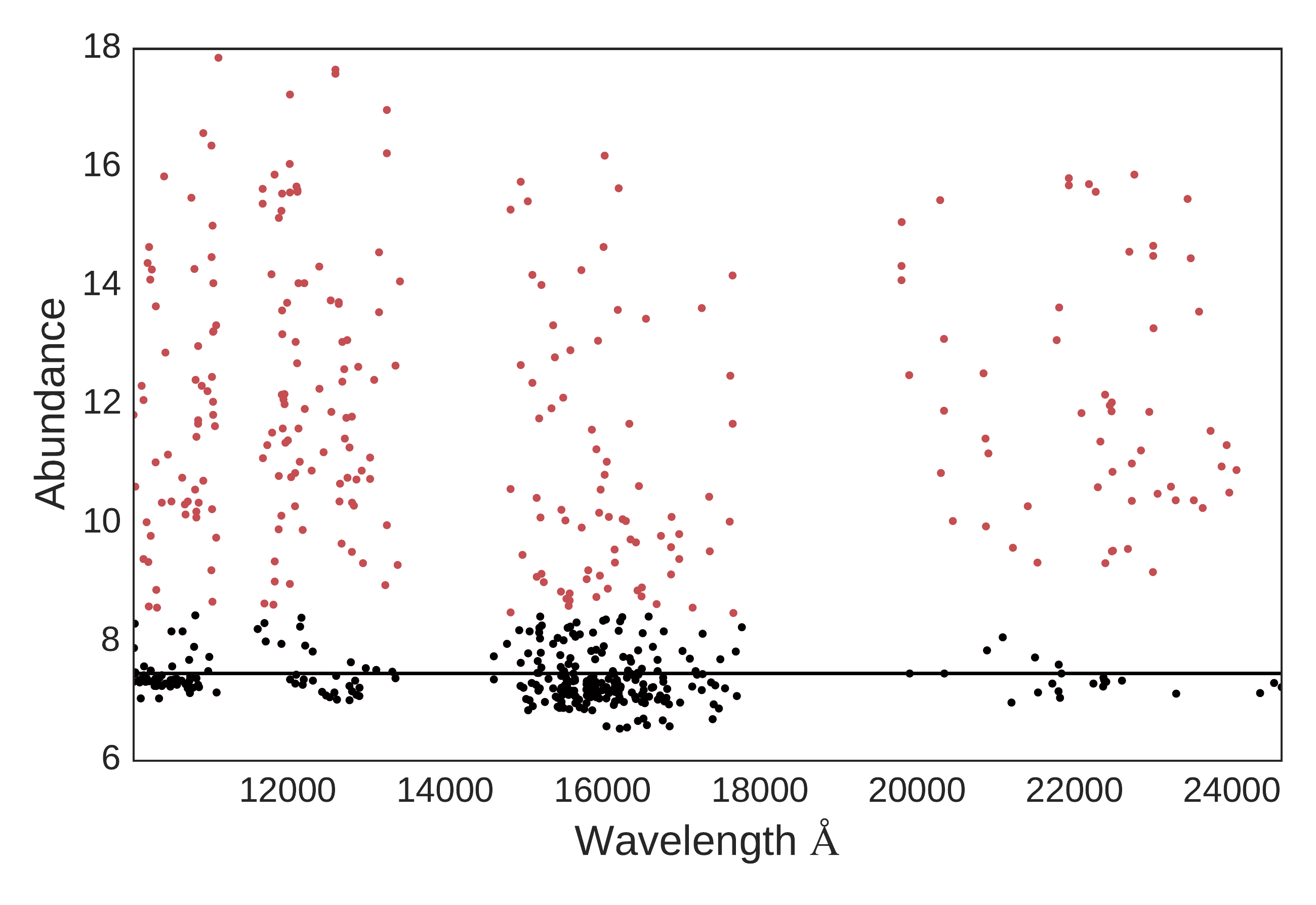}
    \caption{This plot shows the abundances of all \ion{Fe}{i} lines before
    relabration of the $\log\mathit{gf}$ values as a function of the wavelength.
    All red points deviate more than 1 dex from the expected solar value of
    7.47 (horizontal line) and are therefore discarded form the line list.}
    \label{fig:fe1_before_recal}
\end{figure*}

It is clear from Fig.~\ref{fig:fe1_before_recal} that most of the lines taken
from VALD3 has bad $\log\mathit{gf}$ values. This reinforces our need to
use differential analysis also in the NIR.

\end{appendix}

\end{document}